# A versatile scanning photocurrent mapping system to characterize optoelectronic devices based on 2D materials

Christoph Reuter[1,2], Riccardo Frisenda*[2], Der-Yuh Lin[3], Tsung-Shine Ko[3], David Perez de Lara[2] and Andres Castellanos-Gomez*[2,4]

[1] Department of Electrical Engineering and Information Technology, Technische Universität Ilmenau, Gustav-Kirchhoff-Str. 1, Ilmenau 98693, Germany.

[2] Instituto Madrileño de Estudios Avanzados en Nanociencia (IMDEA-nanociencia), Campus de Cantoblanco, E-28049 Madrid, Spain.

[3] National Changhua University of Education, Bao-Shan Campus, No. 2, Shi-Da Rd, Changhua City 500, Taiwan, Republic of China.

[4] Present address: Instituto de Ciencia de Materiales de Madrid (ICMM-CSIC), C/Sor Juana Inés de la Cruz 3, E-28049 Madrid, Spain.

*E-mail: riccardo.frisenda@imdea.org; andres.castellanos@csic.es.

**ABSTRACT:** The investigation of optoelectronic devices based on two-dimensional materials and their heterostructures is a very active area of investigation with both fundamental and applied aspects involved. We present a description of a home-built scanning photocurrent microscope that we have designed and developed to perform electronic transport and optical measurements of two-dimensional materials based devices. The complete system is rather inexpensive (<10000 €) and it can be easily replicated in any laboratory. To illustrate the setup we measure current-voltage characteristics, in dark and under global illumination, of an ultra-thin PN junction formed by the stacking of an n-doped few-layer $MoS_2$ flake onto a p-type $MoS_2$ flake. We then acquire scanning photocurrent maps and by mapping the short circuit current generated in the device under local illumination we find that at zero bias the photocurrent is generated mostly in the region of overlap between the n-type and p-type flakes.

Motivated by their remarkable electronic properties, recently there has been a surge of experimental efforts to apply graphene and other two-dimensional (2D) materials, such as transition metal dichalcogenides (TMDCs), black phosphorous (BP), and others in electronic devices such as field effect transistors, logical circuits, oscillators or flash memories [1-6]. The fabrication of heterostructures based on the stacking of different 2D materials has also shown promising results in the fabrication of rectifiers and more complex devices [7, 8]. Apart from presenting outstanding electronic properties, some of these materials (TMDCs, BP) can interact strongly with light and are considered prospective candidates for novel optoelectronics devices [9, 10]. In fact, in the last years 2D-based detectors with ultrahigh responsivity [11], atomically thin solar cells [12] and ultrafast photodetectors [13] among others have been demonstrated.

In most previous works, the characteristics of 2D-based optoelectronic devices were studied upon wide-field illumination (light spot larger than the device lateral dimensions) which is fast to implement and gives valuable information like the responsivity or the time constant of the device [14, 15]. Nevertheless, given the rich plethora of mechanisms involved in the photocurrent generation process in 2D systems, some physical effects cannot be studied in this kind of measurements. Therefore scanning photocurrent (SPC) measurements, where a small diameter light spot is scanned over the device area to spatially resolve the photocurrent generation have been carried out to better understand the working principles behind 2D based optoelectronic devices [12, 16-21]. Among the different SPC studies found in literature we mention the investigation of the Schottky barriers generated at the interface between TMDCs and metallic electrodes [22-25], photothermoelectric effects in $MoS_2$ [26], the band-offsets at monolayer–multilayer $MoS_2$ junctions [27]. These works have clearly demonstrated how SPC mapping is a very powerful tool in the investigation of optoelectronic devices based on 2D materials. We found, however, that a comprehensive description of the tools and the setup employed to carry out these experiments is missing in the literature (including in PhD dissertations) which is probably hampering the widespread implementation of this useful technique. Moreover, the standard way used in literature to carry out SPC measurements is to employ an optical chopper to modulate the incident light and a lock-in amplifier to record the modulated photocurrent [19, 28]. This measurement scheme has the advantage of allowing the measurement of small electrical signals, for example in samples with a low responsivity and high dark current, given the large achievable signal to noise ratio, while two main disadvantages are the high price of the lock-in amplifier and the large optical table needed to accommodate the necessary components.

In this context we present a thorough description of a home-built scanning photocurrent microscope that we have designed and developed. The complete system is rather inexpensive (<10000 €) and it can be easily replicated (see Table 1 for a full list with all the required components and their part numbers). We illustrate the performance of the present setup by acquiring scanning photocurrent maps in an ultra-thin PN junction formed by the stacking of an n-doped few-layer $MoS_2$ flake onto a p-type $MoS_2$ flake, demonstrating state-of-the-art performances. We thus believe that we provide a simple yet powerful tool that can be implemented in many groups working on the optoelectronic properties of graphene and other 2D materials.

| Part number | Description | Distributor | Price (€) |
|---|---|---|---|
| | Zoom lens tube + Focus stage + PCB camera+ LED illuminator | Hongkong Lapsun Kits (ebay) | 475.00 |
| 716-2040 | Breadboard table, 400 mm x 200 mm | EKSMA optics | 130.00 |
| CCM1-BS 013/M | 50/50 beam splitter cube | Thorlabs | 241.20 |
| SM1A9 | External C-mount to internal SM1 adapter | Thorlabs | 16.88 |
| SM1NR1 | Focusing ring | Thorlabs | 175.50 |
| CXY1 | XY manual adjustment of light spot | Thorlabs | 152.15 |
| M530F2 | Fiber-Coupled 530 nm LED | Thorlabs | 333.00 |
| LEDD1B | T-Cube LED driver | Thorlabs | 263.70 |
| KPS101 | LED driver power supply | Thorlabs | 23.14 |
| M65L01 | 10 µm diameter core fiber | Thorlabs | 99.90 |
| M28L01 | 400 µm diameter core fiber | Thorlabs | 75.69 |
| 960-0070-03LS | Motorized translation stage | EKSMA optics | 2 x 498.00 |
| 980-0942 | 2-axis translational stage controller | EKSMA optics | 940.00 |
| | Keithley 2450 sourcemeter | Keythley | 5000.00 |
| 56-650 | 2X Conversion Lens | Edmund optics | 191.10 |
| RP01/M | Rotation stage | Thorlabs | 82.53 |
| PB1 | Mounting Post Base | Thorlabs | 21.33 |
| P300/M | Ø1.5" Mounting Post, L = 300 mm | Thorlabs | 71.10 |
| RS300/M | Ø25.0 mm Mounting Post, L = 300 mm | Thorlabs | 50.75 |

**Table 1**: Components of the scanning photocurrent setup with indicated the part number, the distributor and the commercial price.

The setup is based on a zoom lens inspection system with coaxial illumination, depicted in Figure 1. A 50:50 beam splitter has been attached to the C-mount camera port of the lens tube. An USB CMOS camera is connected to one of the ports of the beam splitter while the other port is used as an input of the illumination that is going to be focused onto a small diameter spot on the sample, which is brought into the zoom lens through a multimode fiber. The spot illumination port is equipped with a focusing ring that allows one to modify the position of the optical fiber until the fiber core end is exactly placed at the image plane of the lens system. At this condition an image of the circular fiber core end (reduced accordingly to the magnification of the lens) is projected onto the surface of the sample. Therefore one can easily adjust the diameter of the spot illumination by either selecting a multimode fiber with a desired core diameter or by modifying the magnification of the zoom lens system. We address the reader to Figures S1 and S2 of the supporting information for a relationship between the fiber core diameter and the projected illumination spot diameter. The spot illumination port is also supplemented with an xy manual stage that allows one to move the position of the spot in the sample plane, being an useful addition for the initial alignment of the spot in the camera field of view. The source of the spot illumination can be any fiber coupled light source. In our implementation we have chosen high power fiber coupled LEDs because they are inexpensive and safe to use and their light intensity can be manually controlled or modulated up to 5 KHz with an external signal generator.

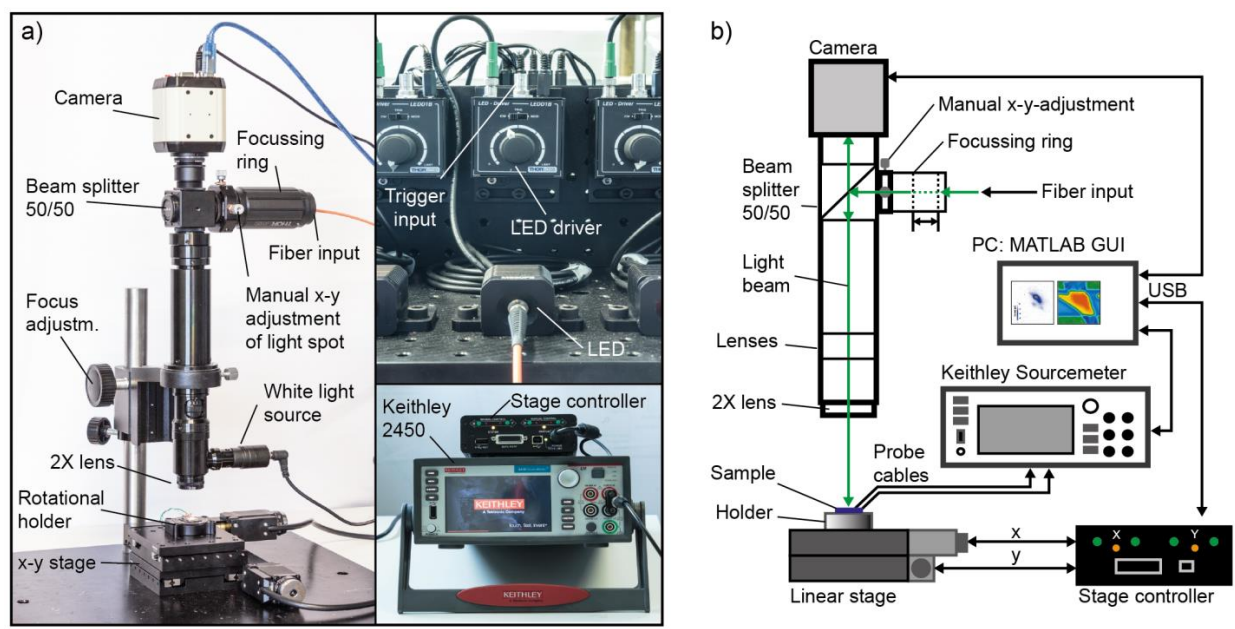

**Figure 1.** a) Pictures of the components of the scanning photocurrent setup with the main components indicated. b) Schematics of the setup circuitry and of the optical path.

The sample is mounted on a rotational sample holder fixed on a motorized xy stage. The two motorized axis have a travel range of 25 mm with a resolution of 1.25 µm in full steps and 0.156 µm in 1/8 steps with a maximum speed of 6 mm/s. The xy stage can be controlled through an USB driver unit, connected to the computer. Whilst scanning the sample, illuminated by a high power LED spot, the device electrical properties are measured with a Keithley 2450 source meter unit as a function of the spot position. The data acquisition and motion control are managed through a home-made routine written in Matlab. A crucial part of the scanning photocurrent measurements is to correlate the photogenerated current with the device geometry. To do so, other SPC mapping systems replace the camera during the scanning by a photodiode. The signal measured by the photodiode, which is proportional to the local reflectivity of the sample, is recorded at each step of the scan, simultaneously to the current, to provide a reflectivity map used to correlate the photocurrent map and the device geometry. In our system we employ directly the signal from the USB camera to construct the reflectivity map. To do so, at each step of the scan we record a snapshot with the camera and we extract the intensity of the spot in the Matlab program. We address the reader to Figures S4 of the supporting information for an example of the reflection map construction.

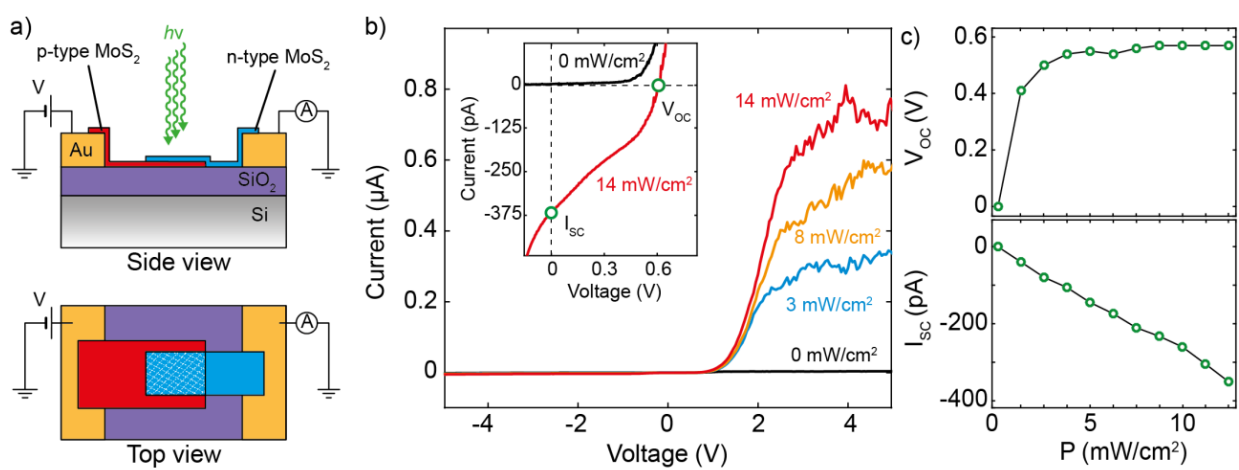

**Figure 2.** a) Schematics of the PN junction based on an n-type doped MoS$_2$ flake stacked on top of a p-type MoS$_2$ flake. b) Current-voltage characteristics of the device in dark conditions (black line) and under external illumination with different power densities (λ = 530 nm). Inset: current-voltage curves at low-bias with indicated the solar cell parameters. c) Open circuit voltage (top) and short circuit current (bottom) as a function of the illumination power density extracted from current-voltage characteristics.

Figure 1a shows pictures of the different components of the SPC system and Figure 1b displays a schematic drawing. In Figures S3 and S5 of the supporting information the reader will find further detailed pictures of the experimental setup. In order to facilitate the implementation of this setup by others Table 1 lists the different components indicating their part number and vendor. To illustrate the performances of this setup we characterize a PN junction, fabricated by stacking n-doped MoS$_2$ flake (MoS$_2$ with 0.5% of Fe substitutional atoms corresponding to a density of approximately $3 \cdot 10^{19}$ atoms/cm$^3$) on top of a p-doped MoS$_2$ flake (0.5% Nb doping) by means of an all dry transfer method schematically depicted in Figure 2a. More details about the fabrication and the characteristics of these PN junction devices can be found in Figures S6 and S7 of the supporting information and in Ref. [29]. From a microscope image of the device we extract the flakes overlap area, A$_{PN}$ = (135 ± 5) μm$^2$.

We first characterize the device under global illumination selecting an optical fiber with a large core (400 μm) that yields a spot on the surface of the sample with a diameter of 33 μm (area A$_{Spot}$ = 3421 μm$^2$ >> A$_{PN}$) that is larger than the device dimensions. The light source is a fiber-couple high power green LED (λ = 530 nm).

Figure 2b displays current-voltage (I-V) characteristics of the device in dark conditions and under increasing illumination power densities (see Figure S8 of the supporting information for a logarithmic representation of the I-Vs). From an inspection of the I-Vs one can see that the device behaves as a diode with a maximum rectification in dark of 250 at 1 V. Upon illumination the forward current increases due to photoconductive effect reaching a saturation current of 0.7 µA at the largest incident power density of 14 mW/cm$^2$. The time dependence of the photocurrent under different bias conditions is shown in Figure S9 of the supporting information. At zero applied bias voltage the built-in potential at the PN junction can effectively separate the photogenerated electron-hole pair giving rise to a photocurrent at zero bias voltage (commonly known as short circuit current, $I_{SC}$). Similarly a voltage builds up across the junction at zero current upon illumination (open circuit voltage, $V_{OC}$). These two parameters can directly be extracted from the I-Vs under illumination as shown in the inset of Figure 2b. Figure 2c shows $I_{SC}$ and $V_{OC}$ plotted as a function of the incident optical power. It can be seen that $I_{SC}$ follows a linear dependency on the power while $V_{OC}$ has a logarithmic dependence, as confirmed by fits, consistent with the classical model of a PN junction. The MoS$_2$ PN junction can be used as a solar cell and Figure S10, in the supporting information, shows the electrical power that can be extracted from the device as a function of the illumination power density. At a power density of 14 mW/cm$^2$, the open circuit voltage is $V_{OC}$ = 0.57 V and the maximum generated power is $P_{MP}$ = 75 pW at a voltage $V_{MP}$ = 0.42 V. From these values we find the figures of merit of the MoS$_2$ PN junction: fill factor F.F. = 0.52 and efficiency η = 0.5%.

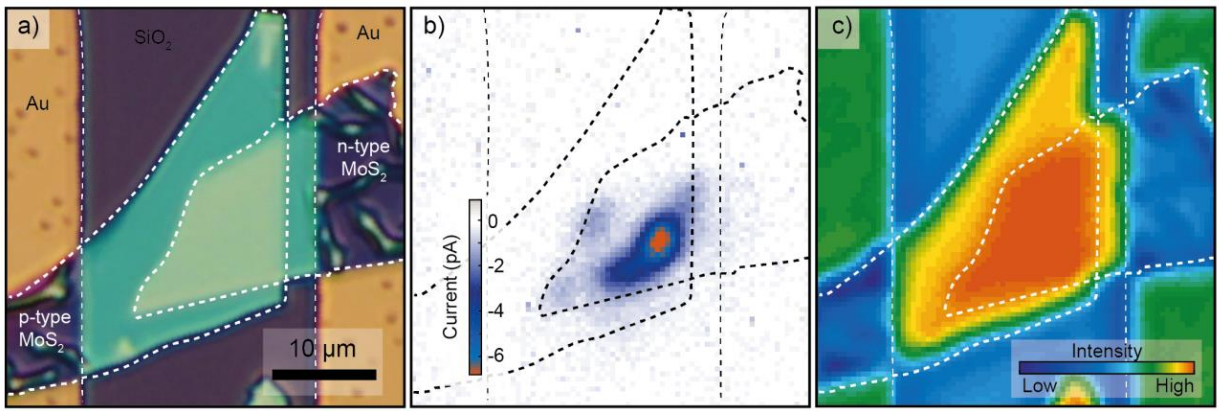

**Figure 3.** a) Optical image of the device. b) Photocurrent image of the MoS$_2$ PN junction with zero bias voltage applied. c) Spatial map of the intensity of the reflected light from the device.

We now turn our attention to SPC maps measurements carried out in the same device. To probe locally the device we replace the 400 µm core multimode optical fiber with a 10 µm core one to produce a spot on the sample surface of (2.3 ± 0.3) µm of diameter, defined as the full width at half maximum of the Gaussian shaped intensity spot profile. This value for the spot-size is approximately two times larger than the theoretical diffraction limited spot size of 1.1 µm given the numerical aperture of our setup N.A. = 0.25. This suggests that additional broadening of the spot is present eventually due to imperfections in the tube lens (see the supporting information). To increase the resolution of the setup one could replace the present lens tube (N.A. = 0.25) with a larger numerical aperture one or to switch to a confocal microscope arrangement. A different approach is to abandon the far-field regime and work in near-field, for example in the near-field scanning microscopy (SNOM) technique [16, 17, 30-32]. Nevertheless, we notice that both these approaches would result in higher setup costs and a more complex experimental arrangement in respect to focusing and alignment than the proposed solution.

Figure 3a shows an optical image of the region of the sample investigated by SPC mapping where one can distinguish the drain and source gold electrodes and the two stacked $MoS_2$ flakes that are partially overlapping. By mapping the photocurrent while the drain-source voltage is kept at zero, one maps the short circuit current $I_{SC}$. The misalignment between the bands of the two differently doped $MoS_2$ flakes generates locally an electric potential that can separate the photoexcited electron hole pairs and give rise to a net current. This process is expected to take place only in the region where the p-type and n-type flakes overlap [12, 21]. During the SPC measurement we block any external light that could hit the sample, apart from the focused laser used to scan.

Figure 3b shows the recorded zero-bias current ($I_{SC}$) map that has been acquired simultaneously to the reflectivity map (Figure 3c) using a step-size of 0.5 μm in both directions. The slow scan axis is parallel to the y-axis while the fast scan axis is parallel to the x-axis in the image. By inspecting the color map of Figure 3b one can see that a (negative) photocurrent, indicated by blue/red color, is generated only in the region of overlap between the two $MoS_2$ flakes (see Figure S11 of the supporting information for a saturated color map that facilitate this observation). The rest of the sample does not display photogeneration of current evidenced by the white color. Notice that the current recorded when the laser spot is located far away from the $MoS_2$ flakes or from the electrodes, corresponds exactly to the dark current of the sample. By subtracting the dark current from the total current recorded in each position one can find the photocurrent generated by the sample.

Interestingly, the photocurrent generation is not homogeneous across the whole overlapping region but instead a hotspot where $I_{SC}$ reaches 6.4 pA is visible. We attribute this spatial inhomogeneous current photogeneration to non-uniform interlayer interaction between the two stacked $MoS_2$ flakes as their interface may contain polymer contaminants or physisorbed adsorbates eventually trapped during the dry-transfer process. The active area of the device $A_D$, identifiable from the photocurrent generation in Figure 3b, appears to be smaller than the overlap area $A_{PN}$. Thus the efficiency calculated in precedence could be underestimated by a factor $A_{PN}/A_D \approx 2$ which, if taken into account, gives an efficiency $\eta = 1\%$. These measurements show the importance of spatially resolved photocurrent maps to deeply understand the performance and the limitations of 2D based optoelectronic devices.

In summary, we presented a scanning photocurrent setup that can be used to map the photoresponse of optoelectronic devices based on 2D materials and that can be operated in a global illumination mode and with local illumination (spot size down to 2 μm). We described the system details and we provided a full list of all the components part numbers to facilitate the implementation of this setup by others. We demonstrate the performance of this setup by mapping the zero-bias photocurrent generated in a n-type $MoS_2$/p-type $MoS_2$ PN junction.

**ACKNOWLEDGEMENTS**

We acknowledge financial support from the European Commission, the MINECO, the Comunidad de Madrid, the Netherlands Organization for Scientific Research (NWO), the German Academic Exchange Service (DAAD) through their RISE-program and from the Ministry of Science and Technology of the Republic of China.

**FUNDING**

A.C.G. European Commission under the Graphene Flagship: contract CNECTICT-604391

A.C.G. MINECO: Ramón y Cajal 2014 program RYC-2014- 01406

A.C.G. MINECO: program MAT2014-58399-JIN

A.C.G. Comunidad de Madrid: MAD2D-CM program (S2013/MIT-3007)

R.F. Netherlands Organization for Scientific Research (NWO): Rubicon 680-50-1515

D.P.dL. MINECO: program FIS2015-67367-C2-1-p

D. Y. and T. S. Ministry of Science and Technology of the Republic of China: MOST 105-2112-M-018-006

D. Y. and T. S. Ministry of Science and Technology of the Republic of China: MOS 105-2221-E-018-025

**COMPETING INTERESTS**

The authors declare no competing financial interests.

# Supplementary Information: A versatile scanning photocurrent mapping system to characterize optoelectronic devices based on 2D materials

**Section 1 – Scanning photocurrent setup**

In the SPC setup the illumination is provided by an optical fiber coupled LED source which is focused onto a circular spot on the sample, which is brought into the zoom lens through a multimode fiber. Figure S1 shows the diameter of the spot focused onto the sample surface measured as a function of the optical fiber core diameter. The insets show actual pictures of the light spot focused onto a $SiO_2$ sample. To have an independent estimation of the spot-size in the case of the smallest optical fiber diameter, we scan a gold structure (thickness 100 nm) on a $SiO_2$/Si substrate, shown in Figure S2a, with the 505 nm LED laser focused on the surface (Figure S2b). The spatially-resolved laser light reflected from the sample (Figure S2c) can be fitted by the convolution of a Heaviside step-function (representing the gold structure profile) and a Gaussian with standard deviation $\sigma = 1$ μm. Figure S3 shows additional pictures of the SPC setup. From the lens working distance and the lens area we can estimate the numerical aperture of the system to be N.A. = 0.25. Figure S4a displays a reflectivity map measured on empty gold electrodes and a line profile taken perpendicular to the electrodes gap is shown in panel (b). Figure S4c shows three actual optical pictures of the light spot reflected from the sample at three different positions along the map. Notice that the spot appears brighter when it is positioned onto the electrodes in respect to the gap.

The microscope and camera have been spatially calibrated by taking pictures of a sample with a 25 μm-period array of squares, shown in Figure S5, and we find that 1 pixel on the camera corresponds to (300 ± 20) nm. To test and calibrate the motion of the xy motorized stage we used a polycarbonate sample with periodical features patterned on top and performed closed loop movements (with the same starting and ending point) like the one shown in Figure S5a. We find that the backlash of the x and y motors is the same and amounts to 1.5 μm as seen from the line profiles in Figure S5b.

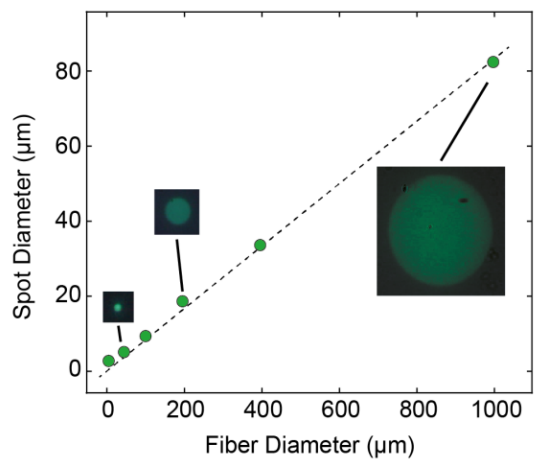

**Figure S1**: Relationship between the size of the fiber core and its image on the sample surface due to the magnification of the optical system. The diameter is the full-width at half-maximum (FWHM) of the light intensity. The inner pictures are microscopic photographs of the light spot focused onto a SiO$_2$/Si substrate.

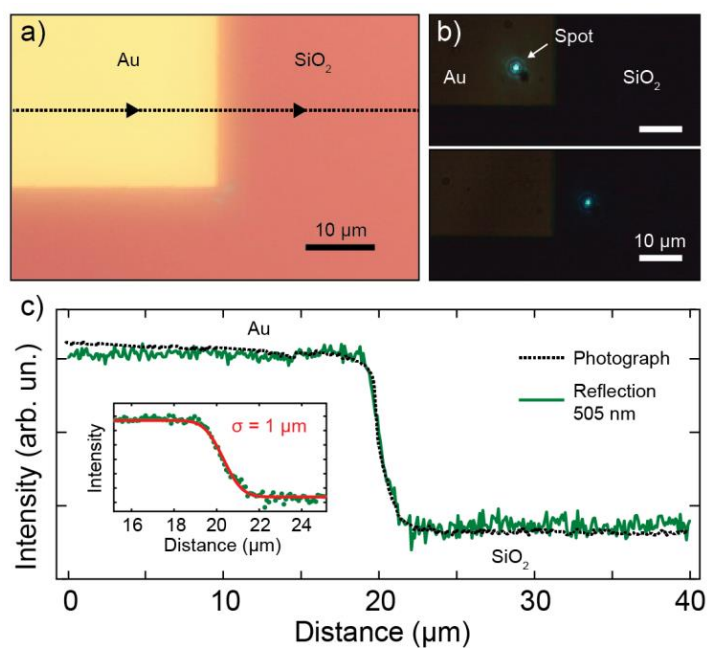

**Figure S2**: a) Microscope image of a gold square (thickness 100 nm) on top of a SiO$_2$/Si substrate. b) Microscope pictures of the same area of panel (a) with the 505 nm laser spot on top of the gold structure (top) and on top of SiO$_2$ (bottom). The laser is scanned along the dashed line drawn in panel (a). c) Line profile extracted from the photograph in (a) and from the reflection of the laser. Inset: The red line is the convolution of a Heaviside step function with a Gaussian from which we estimate the root mean square spotsize σ = 1 μm. At 505 nm of wavelength a diffraction limited system would produce a spotsize with σ = 0.46 µm (≈ 0.42 / (2 · NA)), given a numerical aperture NA ≈ 0.25.

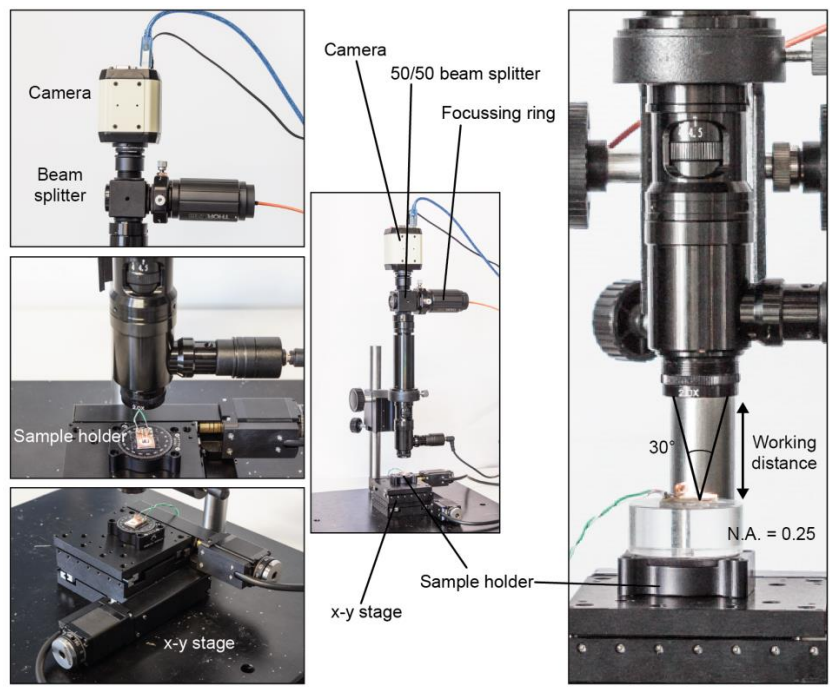

**Figure S3**: Additional pictures of the scanning photocurrent setup.

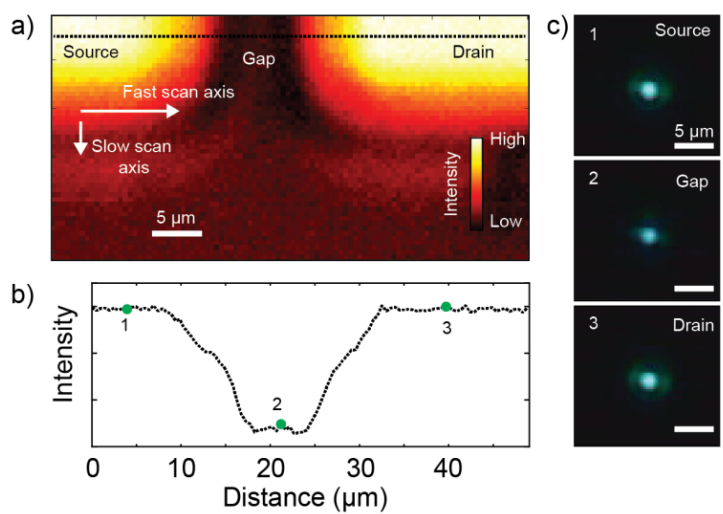

**Figure S4**: a) Spatial map of the intensity of the reflected light from a two-terminal device. b) Line-cut of the intensity at the position indicated by the dashed line in panel (a). c) Optical photographs of the light spot

reflected from the sample at the three positions indicated in panel (b). Notice that he spot appears brighter in positions 1 and 3 when the light is reflected from the metallic electrodes.

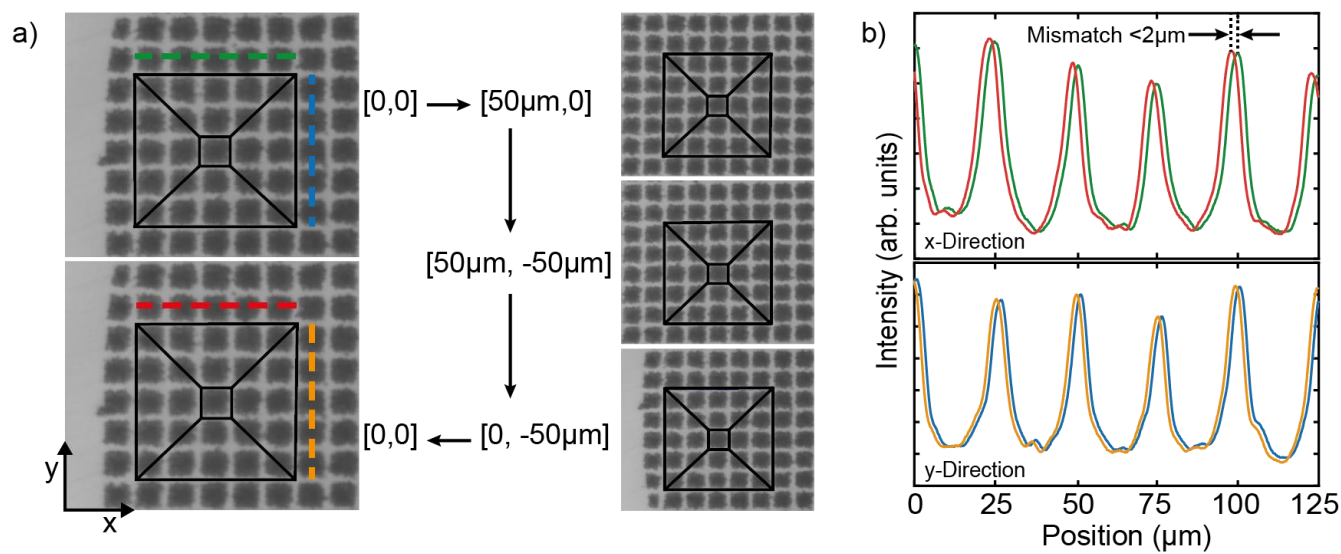

**Figure S5**: a) Characterization of the xy motors movement and the mechanical backlash. b) Intensity line profiles along the lines indicated in panel (a).

### Section 2 – Fabrication of the MoS₂ PN junction

Figure S6 shows the fabrication steps of the p-type $MoS_2$ / n-type $MoS_2$ heterojunction. We first identify a flake of p-type $MoS_2$ onto a PDMS Gel-Pack substrate (first panel, transmission illumination mode optical microscope image). We then deterministically transfer the flake in between two gold electrodes patterned on a $SiO_2$/Si substrate (second panel, epitaxial illumination mode). We repeat the same process for a n-type $MoS_2$ flake (third panel). We transfer the flake onto the p-type flake by carefully aligning the flake in order to have an overlap area between the two flakes (fourth panel, the overlap area is visible as a lighter region). To find the thickness of the flakes we plot in Figure S7 the intensities of the red and blue channel of the microscope transmission mode images of the flakes normalized by the intensity of the PDMS substrate. From a comparison to the transmission intensity dependency on the number of layers shown in Figure S7g, we estimate that the p-type flake thickness is between 8 and 9 layers and that the n-type flake thickness is between 8 and 10 layers.

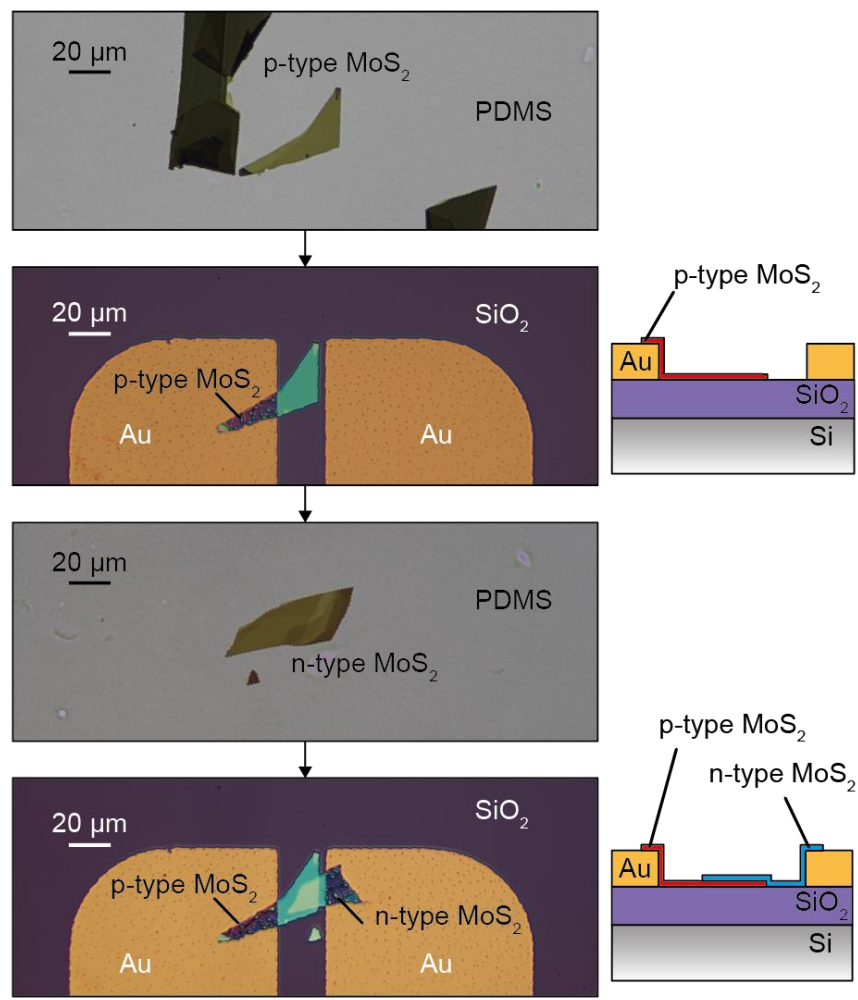

**Figure S6**: Fabrication of a MoS$_2$ PN junction. A p-type MoS$_2$ flake is deposited on PDMS and identified with a microscope. The flake can then be transferred on pre-patterned gold electrodes with all-dry deterministic transfer. Then an n-type flake is identified and finally transferred onto the p-type flake and gold electrodes.

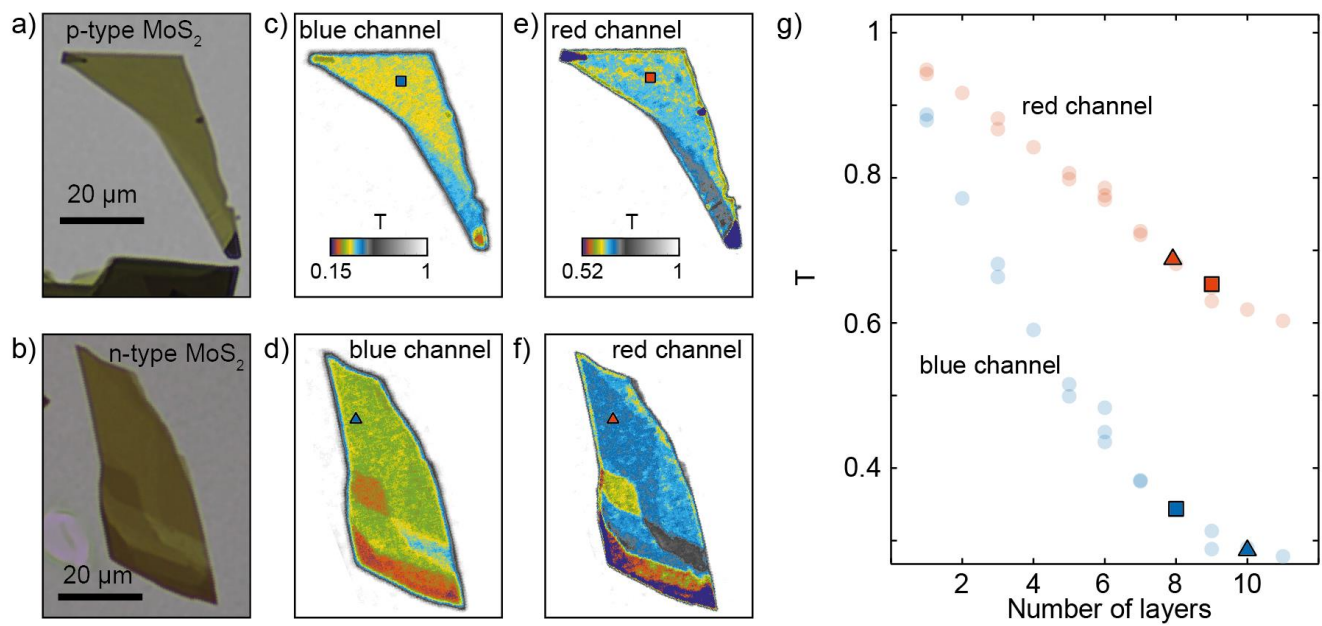

**Figure S7**: a-b) Microscope image of a p-type (n-type) MoS2 flake in transmission illumination mode. c-d) Intensity of the blue channel of the images in (a) and (b) normalized by the intensity of the PDMS substrate (white color that corresponds to unitary transmission is the transmission from the substrate). e-f) Same as (c) and (d) for the red channel, notice that the color scale range is different. g) Transmission of MoS$_2$ flakes as a function of the number of layers. The red and blue square and triangle shapes correspond to the signal of the flakes at the positions indicated in panels (c-f).

## Section 3 – Electrical characterization

Figure S8a shows a logarithmic representation of the current-voltage characteristics in dark and under illumination of the PN junction. Figure S8b shows the rectification ratio of the I-V curve measured in dark calculated as dividing the forward current by the backward current. Figure S9a displays the photocurrent generated in the device at zero bias voltage under 0.1 Hz (square-wave) modulated illumination intensity and Figure S9b shows the time dependency of the photocurrent at positive and negative voltages. The device is stable in time and has a larger response at positive voltages (PN junction forward bias connection) while it responds faster at negative voltages (PN junction reversely bias). In Figure S10 we plot the power output of the device computed from the I-Vs of Figure 2 of the main text. Figure S11 shows the same scanning photocurrent map of the main text with the color-scale saturated in order to highlight the fact that the photocurrent is almost completely generated in the area of overlap between the p-type and n-type flake.

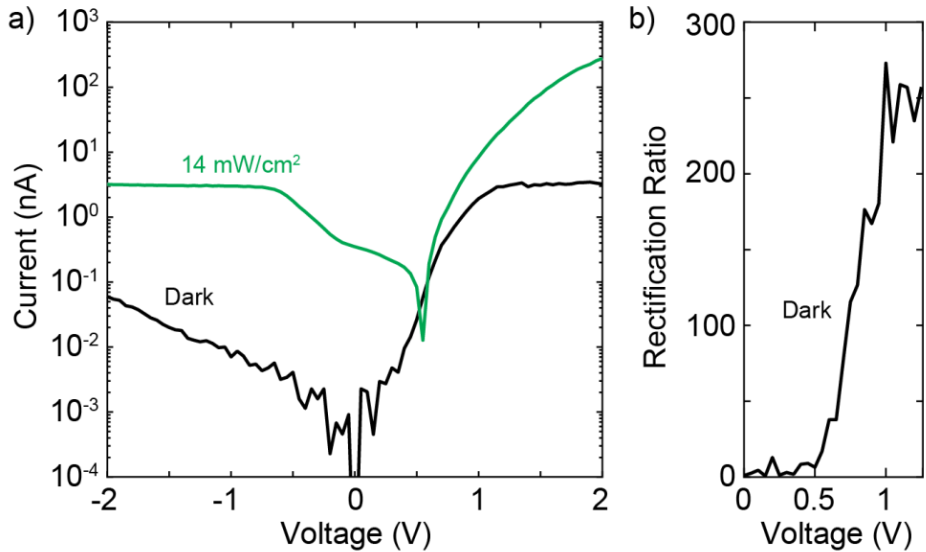

**Figure S8**: a) Current-voltage characteristics of the device in dark conditions (black line) and under external illumination with power density 14 mW/cm$^2$ and λ = 530 nm plotted in semi-logarithmic scale. b) Rectification ratio of the dark current-voltage curve as a function of the applied bias voltage.

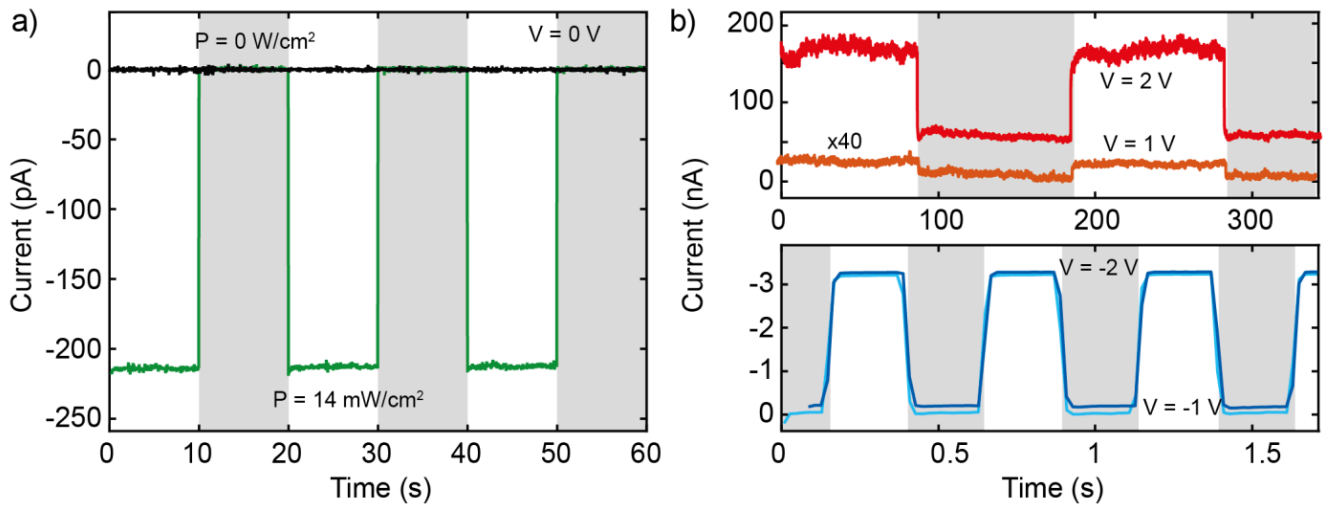

**Figure S9**: a) Time dependency of the PN junction at zero bias voltage with the external light source modulated at 0.1 Hz. b) Time dependency at positive voltage (top) and negative voltage (bottom). Notice that the two graphs have different scales for the time axis.

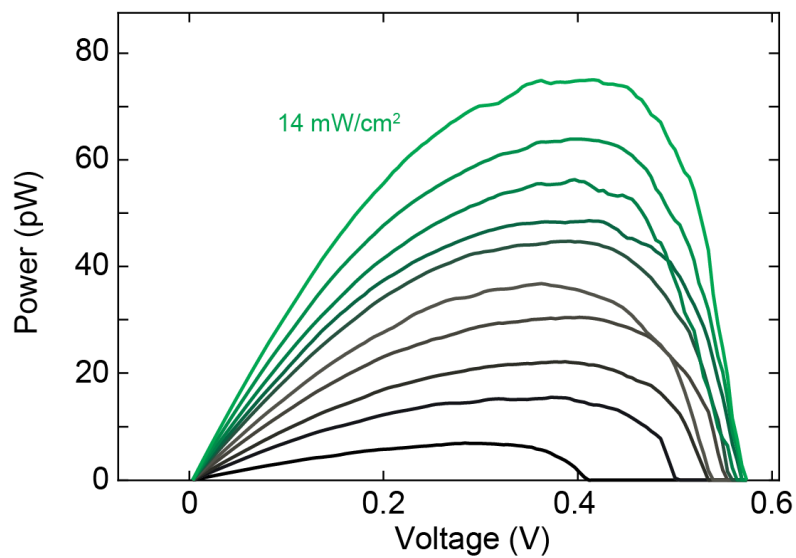

**Figure S10**: Electrical power as a function of bias voltage at different illumination power densities λ = 530 nm.

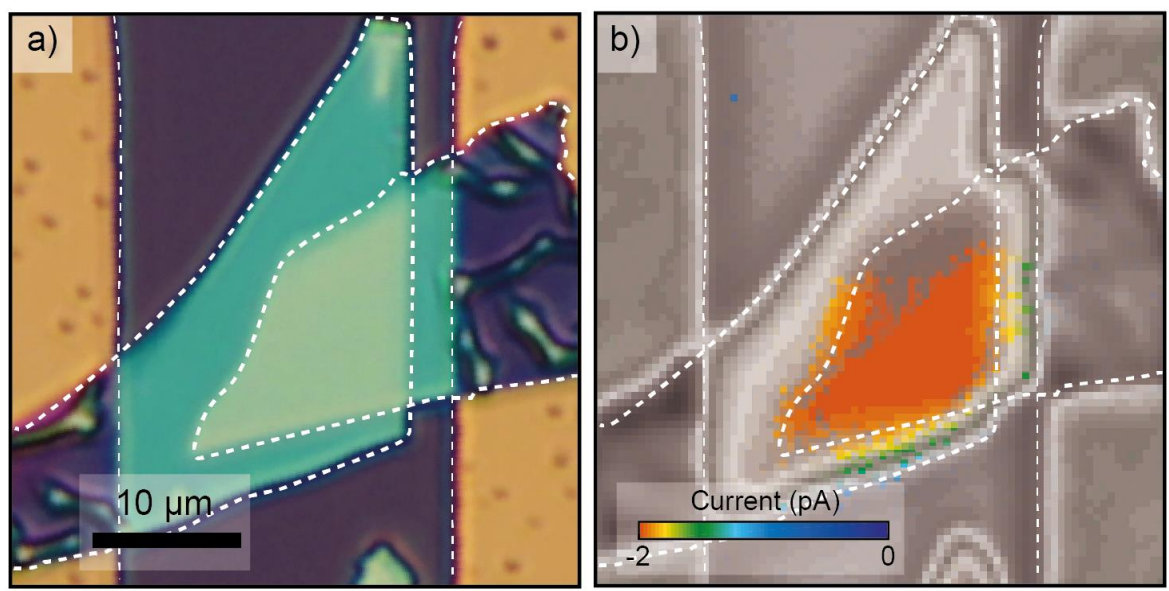

**Figure S11**: a) Optical image of the device. b) Saturated photocurrent map showing clearly that the photocurrent is mainly generated in the p-n overlap region.

**Section 4 – Matlab measurement routine**

In the following we transcribe the matlab routine that we wrote to control the x-y stage and perform the scanning photocurrent measurement. To apply the voltage and read the current we control a Keithley 2450 digital multimeter. The reflection image is compiled by acquiring images with a PCB camera.

```
%%%%%%% Begin set up the matlab graphical interface
 gui_Singleton = 1;
gui_State = struct('gui_Name',      mfilename, ...
    'gui_Singleton',  gui_Singleton, ...
    'gui_OpeningFcn', @Scan_GUI_OpeningFcn, ...
    'gui_OutputFcn',  @Scan_GUI_OutputFcn, ...
    'gui_LayoutFcn',  [] , ...
    'gui_Callback',   []);
if nargin && ischar(varargin{1})
    gui_State.gui_Callback = str2func(varargin{1});
end
 if nargout
    [varargout{1:nargout}] = gui_mainfcn(gui_State, varargin{:});
else
    gui_mainfcn(gui_State, varargin{:});
end
function Scan_GUI_OpeningFcn(hObject, eventdata, handles, varargin)
 set(handles.backlash_edit, 'Enable', 'off');
set(handles.abort_button2,'Enable','off');

global pcm_data;
global reflection_data;
```

```
global selected_cam;

global spot_area;

global xstart;

global xend;

global ystart;

global yend;

selected_cam = 1; %defalt value for the chosen camera

filename = 'temp_data\spot_area.mat';

myVars = {'xstart','xend','ystart','yend'};

spot_area = load(filename,myVars{:});

xstart = spot_area.xstart;

xend = spot_area.xend;

ystart = spot_area.ystart;

yend = spot_area.yend;

axes(handles.axes1);

imagesc(pcm_data);

colorbar;

h = colorbar;

colormap(jet);

xlabel ('X-Steps');

ylabel ('Y-Steps');

ylabel(h, 'Photocurrent [A]');

title('Photocurrent Map');

axes(handles.axes2);

imagesc(reflection_data);
```

```
xlabel ('X-Steps');

ylabel ('Y-Steps');

title('Laser Reflection Image');

drawnow;

pause(0.2);

%%%%%% End set up the matlab graphical interface

 %%%%%% Load .dll for communication with the motor controller

[~,maxArraySize]=computer;

is64bit = maxArraySize > 2^31;

loadlibrary('libximc.dll', @ximcm)

pause(0.2);

%%%%%% Check if Keithley is connected

vinfo = instrhwinfo('visa','NI'); %get info about connected visa devices

obj1a = vinfo.ObjectConstructorName;

obj1 = instrfind('Type', 'visa-usb');

fclose(obj1);
```

```
%%%%%% check connected devices (motors) & name them

device_names = ximc_enumerate_devices_wrap(0);

devices_count = size(device_names,2)

if devices_count == 0

   set(handles.status_edit, 'String',...

      'ERROR: No X-Y-Motors found.');

   return;

end

if devices_count == 1

   set(handles.status_edit, 'String',...

      'ERROR: Only one X-Y-Motor found.');

   return;

end

for i=1:devices_count

   disp(['Found device: ', device_names{1,i}]);

end

device_name1 = device_names{1,1};

device_id1 = calllib('libximc','open_device', device_name1);

device_name2 = device_names{1,2};

device_id2 = calllib('libximc','open_device', device_name2);

%%%%%% Zeroing Motors position
```

```matlab
set(handles.status_edit, 'String', 'Moving to Y-Axis-Homezero...');

result = calllib('libximc','command_homezero', device_id1);

if result ~= 0

    disp(['Homezero1 failed with code', num2str(result)]);

    set(handles.status_edit, 'String', {'Setting Y-Axis-Homezero failed!',...

        'Close MATLAB and check motor connection in XILab Software.'});

end

set(handles.status_edit, 'String', 'Moving to X-Axis-Homezero...');

result = calllib('libximc','command_homezero', device_id2);

if result ~= 0

    disp(['Homezero2 failed with code', num2str(result)]);

    set(handles.status_edit, 'String', {'Setting X-Axis-Homezero failed!',...

        'Close MATLAB and check motor connection in XILab Software.'});

end

set(handles.status_edit, 'String', 'Setting Y-Axis-Homezero as Zero-Position...');

result = calllib('libximc','command_zero', device_id1);

if result ~= 0

    disp(['Zeroing1 failed with code', num2str(result)]);

    set(handles.status_edit, 'String', {'Zeroing Y-Axis failed!',...

        'Close MATLAB and check motor connection in XILab Software.'});

end

set(handles.status_edit, 'String', 'Setting Y-Axis-Homezero as Zero-Position...');

result = calllib('libximc','command_zero', device_id2);
```

```
if result ~= 0

    disp(['Zeroing2 failed with code', num2str(result)]);

    set(handles.status_edit, 'String', {'Zeroing X-Axis failed!',...

        'Close MATLAB and check motor connection in XILab Software.'});

end

pause(1);

set(handles.status_edit, 'String', 'Homezeroing finished!');

result = calllib('libximc','command_wait_for_stop', device_id1, 100);

device_id1_ptr = libpointer('int32Ptr', device_id1);

calllib('libximc','close_device', device_id1_ptr);

result = calllib('libximc','command_wait_for_stop', device_id2, 100);

device_id2_ptr = libpointer('int32Ptr', device_id2);

calllib('libximc','close_device', device_id2_ptr);

%%%%%% Executes on button press in move_button.

function move_button_Callback(hObject, eventdata, handles)

% hObject    handle to move_button (see GCBO)

% eventdata  reserved - to be defined in a future version of MATLAB

% handles    structure with handles and user data (see GUIDATA)

device_names = ximc_enumerate_devices_wrap(0);
```

```
devices_count = size(device_names,2)

if devices_count == 0

    set(handles.status_edit, 'String',...

        'ERROR: No X-Y-Motors found.');

    return;

end

if devices_count == 1

    set(handles.status_edit, 'String',...

        'ERROR: Only one X-Y-Motor found.');

    return;

end

for i=1:devices_count

    disp(['Found device: ', device_names{1,i}]);

end

device_name1 = device_names{1,1};

device_id1 = calllib('libximc','open_device', device_name1);

device_name2 = device_names{1,2};

device_id2 = calllib('libximc','open_device', device_name2);

x_move_string = char(get(handles.move_x_edit,'String'));

y_move_string = char(get(handles.move_y_edit,'String'));

x_move = str2num(x_move_string);

if isempty(x_move_string) %if nothing was entered in GUI -> value=0
```

```
    x_move = 0;

    x_move_string = num2str(x_move);

    set(handles.move_x_edit, 'String', x_move_string);

end

y_move = str2num(y_move_string);

if isempty(y_move_string)

    y_move = 0;

    y_move_string = num2str(y_move);

    set(handles.move_y_edit, 'String', y_move_string);

end

position = ['[',x_move_string,':', y_move_string,']','...'];

position_text = sprintf('Moving to %s%s', position);

set(handles.status_edit, 'String', position_text);

x_pos = fix(x_move / 1.25); %getting just the integer part

r_x = (x_move / 1.25) - x_pos; %getting the remaining floating point part

x_ustep = (r_x * 1.25) / 0.005; % 1.25/256 ~ 0.005µm per microstep

y_pos = fix(y_move / 1.25); %getting just the integer part

r_y = (y_move / 1.25) - y_pos; %getting the floating point par

y_ustep = (r_y * 1.25) / 0.005; % 1.25/256 ~ 0.005µm per microstep

pause(0.2);
```

```
result = calllib('libximc','command_move', device_id2, x_pos, x_ustep);

result = calllib('libximc','command_move', device_id1, y_pos, y_ustep);

result = calllib('libximc','command_wait_for_stop', device_id2, 100);

result = calllib('libximc','command_wait_for_stop', device_id1, 100);

position = ['[',x_move_string,':', y_move_string,']','.'];

position_text = sprintf('Moved to %s%s', position);

set(handles.status_edit, 'String', position_text);

pause(0.2);

result = calllib('libximc','command_wait_for_stop', device_id1, 100);

device_id1_ptr = libpointer('int32Ptr', device_id1);

calllib('libximc','close_device', device_id1_ptr);

result = calllib('libximc','command_wait_for_stop', device_id2, 100);

device_id2_ptr = libpointer('int32Ptr', device_id2);

calllib('libximc','close_device', device_id2_ptr);

%%%%%% Executes on button press in start_button

%%%%%% MAIN PROGRAM ROUTINE

function start_button_Callback(hObject, eventdata, handles)
```

```
handles.stop_now = 0; %Create stop_now in the handles structure, for abort button

guidata(hObject,handles);  %Update the GUI data

set(handles.abort_button2,'Enable','on');

set(handles.status_edit, 'String', 'Preparing...')

set(handles.start_button,'Enable','off');

set(handles.move_button,'Enable','off');

set(handles.move_to_zero_button,'Enable','off');

set(handles.set_zero_button,'Enable','off');

set(handles.set_home_zero_button,'Enable','off');

global pcm_data;

global reflection_data;

global selected_cam;

global spot_area;

global xstart;

global xend;

global ystart;

global yend;

filename = 'temp_data\spot_area.mat'; %load the file with the values for the reflection image

myVars = {'xstart','xend','ystart','yend'};

spot_area = load(filename,myVars{:});

xstart = spot_area.xstart;

xend = spot_area.xend;

ystart = spot_area.ystart;

yend = spot_area.yend;
```

```
cam = selected_cam;

x_stepcounter = 0;

linecounter = 0;

guidata(hObject,handles)

pause(0.2);

%%%%%% Open camera:

vid = videoinput('winvideo', cam);

preview(vid);

if(handles.stop_now) %check if abort button was pushed

    set(handles.abort_button2,'Enable','off');

    set(handles.status_edit, 'String', 'Abort Scan...');

    return;

end

drawnow

handles = guidata(hObject);

%%%%%% get input values for scan parameters

lines = get(handles.nr_of_lines_popup,'Value'); %get # of lines

if lines == 1

    N_lines = 2;

elseif lines == 2
```

```
        N_lines = 4;
elseif lines == 3
        N_lines = 8;
elseif lines == 4
        N_lines = 10;
elseif lines == 5
        N_lines = 16;
elseif lines == 6
        N_lines = 20;
elseif lines == 7
        N_lines = 25;
elseif lines == 8
        N_lines = 32;
elseif lines == 9
        N_lines = 50;
elseif lines == 10
        N_lines = 64;
elseif lines == 11
        N_lines = 75;
elseif lines == 12
        N_lines = 100;
elseif lines == 13
        N_lines = 128;
elseif lines == 14
        N_lines = 150;
elseif lines == 15
        N_lines = 200;
```

```
elseif lines == 16

    N_lines = 256;

end

x_step = str2num(char(get(handles.x_input_edit,'String')));

if isempty(x_step)

    x_step = 1; %default step size

    x_step_string = num2str(x_step);

    set(handles.x_input_edit, 'String', x_step_string);

end

y_step = str2num(char(get(handles.y_input_edit,'String')));

if isempty(y_step)

    y_step = 1; 1; %default step size

    y_step_string = num2str(y_step);

    set(handles.y_input_edit, 'String', y_step_string);

end

N_xsteps = str2num(char(get(handles.N_xsteps_edit,'String')));

if isempty(N_xsteps)

    N_xsteps = N_lines;

    N_xsteps_string = num2str(N_xsteps);

    set(handles.N_xsteps_edit, 'String', N_xsteps_string);

end

N_loops = N_lines; %/ 2; <-for backward and forward movement
```

```
backlash_comp = str2num(char(get(handles.backlash_edit,'String')));

if isempty(backlash_comp)

    backlash_comp = 1.5; %default backl.comp.=1.5µm

    v = num2str(backlash_comp);

    set(handles.backlash_edit, 'String', v);

end

sample_bias = str2num(char(get(handles.sample_bias_edit,'String')));

if isempty(sample_bias)

    sample_bias = 0; %set default bias voltage to 0V

    v = num2str(sample_bias);

    set(handles.sample_bias_edit, 'String', v);

end

current_limit = str2num(char(get(handles.current_limit_edit,'String')));

if isempty(current_limit)

    current_limit = 1; %set default current limit to 1mA

    currentLim_mA_copy = current_limit;

    w = num2str(currentLim_mA_copy);

    set(handles.current_limit_edit, 'String', w);

end

v = num2str(sample_bias);

currentLim_mA = current_limit/1000;

w = num2str(currentLim_mA);

pause(0.2);

bias_string = [':SOURce:VOLT ', v];
```

```
currentLim_string = [':SOURce:VOLT:ILIM ', w];

N_avg = str2num(char(get(handles.averaging_edit,'String')));

if isempty(N_avg)

    N_avg = 1;

    N_avg_copy = N_avg;

    N_avg_string = num2str(N_avg_copy);

    set(handles.averaging_edit, 'String', N_avg_string);

end

time_break = str2num(char(get(handles.time_step_edit,'String')));

if isempty(time_break)

    time_break = 0; %set zero delay between each step

    time_break_copy = time_break;

    time_break_string = num2str(time_break_copy);

    set(handles.time_step_edit, 'String', time_break_string);

end

%%%%%%

handles.stop_now = 0; %Create stop_now in the handles structure

guidata(hObject,handles);  %Update the GUI data

drawnow;

% Configure instrument object, obj1 Keithley

set(obj1, 'InputBufferSize', 10000);
```

```matlab
% Configure instrument object, obj1

set(obj1, 'OutputBufferSize', 512);

fopen(obj1);

fprintf(obj1, '*RST'); %Reset

fprintf(obj1, ':TRACe:MAKE "ScanBuffer", 100000');

fprintf(obj1, ':SOURce:FUNCtion VOLT'); %Sourcing voltage

fprintf(obj1, bias_string); %setting voltage

fprintf(obj1, currentLim_string); %setting current limit

fprintf(obj1, ':OUTPut ON');

fprintf(obj1, ':SENSe:FUNCtion "CURRent"');

Irange=10E-9;

Irange_string = num2str(Irange);

Irange_out = [':SENSe:CURRent:RANGe ', Irange_string];

fprintf(obj1, '%s', Irange_out);

%%%%%%

%%%%%% Main scanning loop

%%%%%%
```

```matlab
set(handles.status_edit, 'String', 'Scanning...');

disp('Scanning...');

img=0;

for i=1:1:N_loops

    if(handles.stop_now) %check if abort button was pushed

        set(handles.abort_button2,'Enable','off');

        set(handles.status_edit, 'String', 'Abort Scan...');

        break;

    end

    drawnow; %refresh GUI

    handles = guidata(hObject);

    %%%%%% step in fast axis direction

    linecounter = linecounter + 1;

    scan_progress = sprintf('Current line: %d of %d', linecounter, N_lines);

    set(handles.status_edit, 'String', {'Scanning...', scan_progress});

    pause(0.1);

    %inner loop for line in x-direction:

    for n=1:N_xsteps

        x_stepcounter = x_stepcounter + 1;

        x_move = x_step * x_stepcounter;

        x_pos = fix(x_move/1.25);
```

```
r_x = (x_move/1.25) - x_pos;

x_ustep = (r_x*1.25)/0.005;

pause(0.01);

avg_array = zeros(N_avg, 1);

for avg=1:N_avg
    fprintf(obj1, ':READ? "ScanBuffer", READ');
    current_string = fscanf(obj1);
    current_num = str2double(current_string);
    avg_array(avg,1) = [current_num];
end

cam_img = getsnapshot(vid); %get camera image

current_avg = sum(avg_array)/N_avg; %final averaging

line = (N_lines - linecounter) + 1; %calculate y-adress in data array

pcm_data(line, n) = [current_avg];

image_gray = rgb2gray(cam_img); %convert camera img to grayscale

image_gray=double(cam_img(:,:,:)); % RICCARDO

%extract zoomed in region from image:

image_zoom = image_gray(ystart:yend,xstart:xend);

intensity = sum(sum(sum(image_zoom)));

reflection_data(line, n) = [intensity];
```

```
x_movement = calllib('libximc','command_move', device_id2, x_pos, x_ustep);

x_stopped = calllib('libximc','command_wait_for_stop', device_id2, 100);

pause(time_break); %programmable break between each step

if(handles.stop_now) %check if abort button was pushed

    set(handles.abort_button2,'Enable','off');

    set(handles.status_edit, 'String', 'Abort Scan...');

    break;

end

handles = guidata(hObject);

drawnow; %updating GUI

end

%%%%%% step in slow axis direction

%java.lang.System.gc(); %Call Java garbage collection, to maintain speed

x_stepcounter = 0;

y_move = y_step * linecounter;

y_pos = fix(y_move/1.25);

r_y = (y_move/1.25) - y_pos;

y_ustep = (r_y*1.25)/0.005;
```

```
pause(0.2);

result = calllib('libximc','command_move', device_id1, y_pos, y_ustep);

result = calllib('libximc','command_wait_for_stop', device_id1, 100);

if(handles.stop_now)

    set(handles.abort_button2,'Enable','off');

    set(handles.status_edit, 'String', 'Abort Scan...');

    break;

end

axes(handles.axes1);

imagesc(pcm_data*1e12);

colorbar;

h = colorbar;

colormap(jet);

title('Photocurrent Map');

xlabel('X-Steps');

ylabel('Y-Steps');

ylabel(h, 'Photocurrent [pA]');

axes(handles.axes2);

imagesc(reflection_data);

title('Laser Reflection Image');

xlabel('X-Steps');

ylabel('Y-Steps');
```

```
    guidata(hObject,handles);

    drawnow;

    handles = guidata(hObject);

end

fprintf(obj1, ':OUTPut OFF');

%%%%%%

%%%%%% End of main scanning loop

%%%%%%

filename = '\temp_data\pcm_data.xlsx';

xlswrite(filename,pcm_data);

filename = '\temp_data\reflection_data.xlsx';

xlswrite(filename,reflection_data);

if(handles.stop_now)

    set(handles.status_edit, 'String', 'Scan aborted!');

end

closepreview(vid);

set(handles.abort_button2,'Enable','off');
```

```
set(handles.move_button,'Enable','on');

set(handles.move_to_zero_button,'Enable','on');

set(handles.set_zero_button,'Enable','on');

set(handles.set_home_zero_button,'Enable','on');

set(handles.start_button,'Enable','on');

%current_folder = pwd; %get current directory

%export_folder = strcat(current_folder,'\temp_data\screenshot.png');

%export_fig (export_folder, '-png', '-transparent'); %write screenshot

set(handles.status_edit, 'String', {'Scan finished!'});

instrreset

disp('Scan finished !');

%%%%%% Executes on button press in save_data_button.

function save_data_button_Callback(hObject, eventdata, handles)

global pcm_data;

[filename, folder] = uiputfile('*.xlsx', 'Save scan data to');

outname = fullfile(folder, filename);

if filename == 0

    return;

end

xlswrite(outname,pcm_data);
```

```
 %%%%% Executes on button press in save_lr_data_button.

function save_lr_data_button_Callback(hObject, eventdata, handles)

global reflection_data;

[filename, folder] = uiputfile('*.xlsx', 'Save scan data to');

outname = fullfile(folder, filename);

if filename == 0

    return;

end

xlswrite(outname,reflection_data);
```